\documentclass[11pt]{article}
\usepackage{amssymb}
\usepackage{latexsym}

\newcommand{\be}{\beta}
\newcommand{\pa}{\partial}

\begin{document}

\begin{titlepage}

 \rightline{Preprint LPT-ORSAY 00-90}

\vskip 1.6cm

\begin{center}

{\Large Exact Solvability of Superintegrable Systems}

\vskip 0.8cm

 Piergiulio Tempesta \cite{byline},\\
 Dipartimento di Fisica, Università  di Lecce and INFN sezione di Lecce, 73100 Italy\\[10pt]

Alexander V. Turbiner  \cite{byline1},\\
 Laboratoire de Physique Theorique, Universit\'e Paris-Sud, France and
 Instituto de Ciencias Nucleares, UNAM, A.P. 70-543, 04510 M\'exico\\[10pt]

 and\\[10pt]
 Pavel Winternitz \cite{byline3},\\
Centre de Recherches Math\'ematiques, Universit\'e de Montreal,  C.P. 6128,
succ. Centre-ville, Montr\'eal (QC) H3C 3J7, Canada

\vskip 0.8cm

\end{center}

\vskip 2.cm

\centerline{Abstract}

\begin{quote}
\hskip .5cm
 It is shown that all four superintegrable quantum systems on the
 Euclidean plane possess the same underlying hidden
 algebra $sl(3)$. The gauge-rotated Hamiltonians, as well as their
 integrals of motion, once rewritten in appropriate coordinates, preserve
 a flag of polynomials. This flag corresponds to highest-weight
 finite-dimensional representations of the $sl(3)$-algebra, realized
 by first order differential operators.
\end{quote}

\end{titlepage}

\section{Introduction}

The purpose of this Letter is to establish a relation between two different
concepts in quantum mechanics: \textit{``superintegrability''} and \textit{%
``exact solvability''}. More specifically we relate these two concepts in
nonrelativistic quantum mechanics in two dimensional Euclidean space $E_{2}$.

The notion of integrability in quantum mechanics \cite{Dirac:1981} comes
naturally as a generalization of a similar notion in classical mechanics
(see, for instance, \cite{Goldstein:1980}). A quantum mechanical system in $%
E_n$ described by the stationary Schroedinger equation
\begin{equation}  \label{e.1}
H \Psi = E \Psi\ ,\ H=-\frac{1}{2}\triangle+V(x_1,,\ldots, x_n)\ ,
\end{equation}
is completely integrable if there exists a set of $(n-1)$ algebraically
independent linear operators $X_a, a=1,2,\ldots,n-1$ commuting with the
Hamiltonian and among each other
\begin{equation}  \label{e.2}
[H, X_a]=0\ ,\ [X_a, X_b]=0\ .
\end{equation}
The system is \textit{``superintegrable"} if there exist $k$ additional
operators, $Y_b, b=1,\ldots, k$, where $0< k \leq (n-1)$, commuting with the
Hamiltonian. It is ``maximally superintegrable'' if $k=n-1$.

The operators $X_{a},Y_{b}$ are usually assumed to be polynomials in the
momenta $\{p_{1},\ldots ,p_{n}\}$ with coordinate dependent coefficients. A
systematic search for superintegrable systems in $E_{2}$ and $E_{3}$ was
conducted some time ago \cite{Fris:1965, Wint:1966, Mak:1967}. A restriction
was imposed, namely that the operators $X_{a}$ and $Y_{b}$ should be second
order polynomials in momenta. It turned out that the existence of such
commuting operators leads to the separation of variables in the Schroedinger
equation. Superintegrable systems are actually \textit{''superseparable''}:
they allow the separation of variables in at least two coordinate systems.

A large body of literature exists on superintegrable systems in $E_n$ \cite
{Fris:1965} - \cite{Sheftel:2000}. In particular, it has been shown recently
\cite{Sheftel:2000} that the superintegrable systems in $E_2$ are
characterized by the existence of at least two generalized Lie symmetries.

Quantum mechanical problems which can be called \textit{exactly solvable}
are defined quite differently. They are characterized by the fact that one
can indicate explicitly an infinite flag of functional linear spaces, which
is preserved by the Hamiltonian \cite{Turbiner:1994}. We recall that a
\textit{flag} is formed by an infinite set of functional linear spaces which
can be ordered in such a way that each of them contains the previous one as
a subspace. One important particular example of these flags is given by
finite-dimensional representation spaces of semi-simple Lie algebras of
first order differential operators. In this case the Hamiltonian is an
element of the universal enveloping algebra of a Lie algebra.

In order to clarify the situation let us consider as an example the case of
one dimensional (quasi)-exactly solvable problems\cite{Turbiner:1988}. Due
to Sophus~Lie it is known that the only Lie algebra of first order
differential operators which acts on the real line and possesses
finite-dimensional representations is the $sl(2,\mathbb{R})$-algebra (for a
discussion see, for example, \cite{Olver:1991, Gonzalez:1992}), realized as
\begin{equation}
J^+_n = x^2 d_x - n x\ \ ,\ \ J^0_n = x d_x - {\frac{n }{2}}\ \ ,\ \ J^-_n =
d_x \ .  \label{sl2}
\end{equation}
For integer $n$ the generators (\ref{sl2}) possess a common invariant
subspace $\mathcal{P}_n=\langle x^k|0\leq k \leq n \rangle$, which is the
linear space of polynomials. It is evident that the spaces $\mathcal{P}_n$
as functions of the parameter $n$ form a flag. This flag is preserved by any
element of the universal enveloping algebra of the $sl(2,\mathbb{R})$
parabolic subalgebra $J^0_n, J^-_n$ for any $n$. Therefore, an element of
this enveloping algebra can be viewed as a Hamiltonian, which defines an
\textit{exactly solvable} system. In a similar manner one can introduce the
notion of a quasi-exactly solvable problem for which the Hamiltonian
possesses the invariant subspace $\mathcal{P}_n$. It can be proven \cite
{Turbiner:1992} that a necessary and sufficient condition for a one
dimensional Hamiltonian to be quasi-exactly solvable is that it belongs to
the universal enveloping algebra of the $sl(2,\mathbb{R})$-algebra taken in
realization (\ref{sl2}).

For two-dimensional problems there exist four candidates for
underlying hidden Lie algebra \cite{Shifman:1989, Olver:1991,
Gonzalez:1992, Turbiner:1994}: $sl(3,%
\mathbb{R})$, $sl(2,\mathbb{R}) \oplus sl(2,\mathbb{R}), o(3,1)$, a
parametric family $gl(2,\mathbb{R})\ltimes {\mathbf{R}}^{r+1}$ and some of
their subalgebras. In particular, the algebra $gl(3,\mathbb{R}) \supset sl(3,%
\mathbb{R})$, realized as
\[
J_1\ =\ \partial_t \ ,\ J_2\ =\ \partial_u \ ,\
\]
\begin{equation}  \label{sl3}
J_3\ =\ t\partial_t \ ,\ J_4\ =\ u\partial_u \ ,\ J_5\ =\ u\partial_t \ ,\
J_6\ =\ t\partial_u \ ,
\end{equation}
\[
J_7\ =\ t^2 \partial_t \ + \ t u \partial_u\ - \ n t \ ,\ J_8\ =\ t u
\partial_t \ + \ u^2 \partial_u\ - \ n u\ ,
\]
\[
X\ =\ n\ ,
\]
will be used in this article as well as the maximal parabolic subalgebra of $%
sl(3,\mathbb{R})$ formed by the generators $\{J_{1,...,6}\}$.

For integer $n$ the generators (\ref{sl3}) possess a common invariant
subspace $\mathcal{P}_n^{(2)}=\langle t^k u^m |0\leq k+m \leq n \rangle$,
which is the linear space of polynomials. Similarly to the case of $sl(2,%
\mathbb{R})$ the spaces $\mathcal{P}_n^{(2)}$ as functions of the parameter $%
n$ form a flag. This flag is preserved by any element of the universal
enveloping algebra of the parabolic subalgebra $\{J_{1,...,6}\}$ \cite
{Turbiner:1994}. Therefore, an element of this enveloping algebra viewed as
a Hamiltonian defines an \textit{exactly solvable} system with the hidden
algebra $sl(3,\mathbb{R})$.

Both superintegrable and exactly solvable systems have numerous applications
in physics. Among the simplest superintegrable ones are the Coulomb system
and the harmonic oscillator in spaces of any dimension. The celebrated
many-body Calogero model is superintegrable as well as the Hartmann
potential of quantum chemistry \cite{Kibler, Hartmann}.

\section{ Hidden Algebra of Superintegrable Hamiltonians}

There exist precisely four quantum (and also classical) Hamiltonians defined
on $E_2$, characterized by two integrals of motion, $[H,X_{1,2}]=0$, such
that $X_{1,2}$ are quadratic in the momenta \cite{Fris:1965,
Wint:1966,Sheftel:2000}. Thus, they are maximally superintegrable and for
their classical counterparts all trajectories are closed. It was shown that
these four Hamiltonians exhaust the list of two-dimensional Hamiltonians
characterized by two integrals of motion in the form of second order
differential operators. They admit the separation of variables in two (or
even more) different coordinate systems. We will show that all of them
possess a hidden $sl(3,\mathbb{R})$ algebra. In particular, this implies
that there exists a coordinate system where the Hamiltonian, as well as the
integrals of motion, after a similarity transformation (gauge rotation) can
be rewritten in terms of the generators of the maximal parabolic subalgebra
of $sl(3,\mathbb{R})$. Furthermore, two of these Hamiltonians have a
striking feature. For each of them, multiplied by a suitable factor $f$,
there exists another set of two commuting operators, $[fH, Y_{1,2}]=0$. This
can be considered as a generalization of the notion of integrability in
quantum mechanics: commuting operators appear not for the Hamiltonian, but
for the Hamiltonian multiplied by a factor.

\textbf{Case I.} The first Hamiltonian written in Cartesian corrdinates is
given by
\begin{equation}  \label{H.1}
H_I(x,y;\frac{\omega^2}{2},\frac{A}{2},\frac{B}{2}) = - \frac{1}{2}%
(\partial^2_x + \partial^2_y) + \frac{\omega^2}{2} (x^2+y^2) + \frac{1}{2}%
\bigg(\frac{A}{x^2} + \frac{B}{y^2}\bigg)\ ,
\end{equation}
where $A, B > - 1/8$ are parameters. The corresponding Schroedinger equation
separates in three different coordinate systems: Cartesian, polar and
elliptical. The eigenfunctions can be written in the form
\begin{equation}  \label{H.2}
\Psi_{n,m}(x,y)=x^{p_1} y^{p_2} L^{(-1/2+p_1)}_n (\omega x^2)
L^{(-1/2+p_2)}_m(\omega y^2) e^{-\frac{\omega x^2}{2}-\frac{\omega y^2}{2}}\
,
\end{equation}
where $L_k^{(\alpha)}(z)$ are Laguerre polynomials, $n,m = 0,1,2, \ldots$
and the parameters $p_{1,2}$ are defined by $A =p_1(p_1-1), B = p_2(p_2-1)$.
The degree of degeneracy of eigenstates is given by a number of partitions
of an integer into the sum of two integers.

We perform a gauge rotation of $H_I (x,y)$, using the ground state
eigenfunction $\Psi_{0,0}(x,y)$ as a gauge factor and then a change of
variables
\[
h^{I} \equiv \frac{1}{\omega} ( \Psi_{0,0}(x,y))^{-1} H_I (x,y)
\Psi_{0,0}(x,y)|_{t=\omega x^2,u=\omega y^2} =
\]
\begin{equation}  \label{HI.1}
-2t\partial^2_t - 2u\partial^2_u + 2t\partial_t + 2u \partial_u - (2p_1+1)
\partial_t - (2p_2+1) \partial_u +1+p_1+p_2\ ,
\end{equation}
with eigenvalues $E(n,m)=n+m,\, n,m = 0,1,2,\ldots$.

It is easy to check that after a gauge rotation with the same gauge factor $%
\Psi_{0,0}(x,y)$ and a change of variables for the integrals of motion, we
arrive at the operators
\begin{equation}
{\hat x}_C^I= 2t \partial^2_t - 2u \partial^2_u - 2t\partial_t + 2u
\partial_u + (2p_1+1) \partial_t - (2p_2+1) \partial_u - p_1 + p_2\ ,
\end{equation}

\begin{equation}
{\hat{x}}_{R}^{I}\ = 4tu(\partial _{t}-\partial
_{u})^2+ 2 [(2p_{1}+1)u-(2p_{2}+1)t] (\partial _{t}-\partial
_{u})-(p_{1}+p_{2})^{2}\ .
\end{equation}
These three operators $h^{I},{\hat{x}}_{C}^{I},{\hat{x}}_{R}^{I}$ obey the
commutation relations
\[
\lbrack h^{I},{\hat{x}}_{C}^{I}]\ =\ [h^{I},{\hat{x}}_{R}^{I}]\ =\ 0\ ,
\]
and
\[
\lbrack {\hat{x}}_{C}^{I},{\hat{x}}_{R}^{I}]=32tu\partial
_{tuu}^{3}-32tu\partial _{ttu}^{3}-8t(2p_{2}+1-2u)\partial
_{tt}^{2}+8u(2p_{1}+1-2t)\partial _{uu}^{2}+
\]
\begin{equation}
+16[(2p_{2}+1)t-(2p_{1}+1)u]\partial
_{tu}^{2}-4(2p_{1}+1)(2p_{2}+1-2u)\partial
_{t}+4(2p_{2}+1)(2p_{1}+1-2t)\partial _{u}  \label{I.0}
\end{equation}
They generate an infinite-dimensional algebra.

The operators $h^{I},{\hat{x}}_{C}^{I},{\hat{x}}_{R}^{I}$ as well as the
commutator $[{\hat{x}}_{C}^{I},{\hat{x}}_{R}^{I}]$ can be immediately
rewritten in terms of the generators $\{J_{1,...,6}\}$ of the maximal
parabolic subalgebra of $sl(3,\mathbb{R})$. They have the form
\begin{equation}
h^{I}\ =\
-2J_{3}J_{1}-2J_{4}J_{2}+2J_{3}+2J_{4}-(2p_{1}+1)J_{1}-(2p_{2}+1)J_{2}\ ,
\label{I.1}
\end{equation}
\begin{equation}
{\hat{x}}_{C}^{I}\ =\
2J_{3}J_{1}-2J_{4}J_{2}-2J_{3}+2J_{4}+(2p_{1}+1)J_{1}-(2p_{2}+1)J_{2}\ ,
\label{I.2}
\end{equation}
\[
{\hat{x}}_{R}^{I}\ =\ 4J_{3}J_{5}+4J_{4}J_{6}-8J_{3}J_{4}+
\]
\begin{equation}
2(2p_{1}+1)J_{5}-2(2p_{2}+1)J_{3}-2(2p_{1}+1)J_{4}+2(2p_{2}+1)J_{6}\ .
\label{I.3}
\end{equation}
The commutation relation (\ref{I.0}) is rewritten as
\[
\lbrack {\hat{x}}_{C}^{I},{\hat{x}}%
_{R}^{I}]=32J_{4}J_{3}(J_{2}-J_{1})-16J_{4}J_{6}+16J_{3}J_{5}
\]
\[
+8(2p_{1}+1)J_{4}(J_{2}-2J_{1})+8(2p_{2}+1)J_{3}(2J_{2}-J_{1})
\]
\begin{equation}
+8(2p_{1}+1)J_{5}-8(2p_{2}+1)J_{6}+4(2p_{1}+1)(2p_{2}+1)(J_{2}-J_{1})\ .
\label{I.4}
\end{equation}

Evidently, the operators (\ref{I.1})-(\ref{I.4}) preserve a triangular flag
of polynomials $\mathcal{P}^{(2)}$ in $t,u$:
\[
h(t): \mathcal{P}_{n}(t,u)\mapsto \mathcal{P}_{n}(t,u)
\]
where $\mathcal{P}_{n}(t,u)=\langle t^p u^q|0\leq p+q \leq n\rangle$. Hence,
the operators $h^{I},{\hat x}_C^I,{\hat x}_R^I$ are characterized by
infinitely many finite-dimensional invariant subspaces and thus possess
infinitely many polynomial eigenfunctions.

\textbf{Case II.}

The second superintegrable Hamiltonian can be separated in Cartesian and
parabolic coordinates. In Cartesian coordinates it is given by
\begin{equation}  \label{HII.1}
H_{II}(x,y) = - \frac{1}{2}(\partial^2_x + \partial^2_y) + 2 \omega^2 x^2 +
\frac{\omega^2}{2} y^2 + \frac{B}{2 y^2}\ ,
\end{equation}
where $B > - 1/8$ is a parameter. The eigenfunctions and eigenvalues have
the form
\begin{equation}  \label{HII.2}
\Psi_{n,m}(x,y)= y^{p_2} H_n (\sqrt{2\omega} x) L^{(-1/2+p_2)}_m(\omega y^2)
e^{- \omega x^2 -\frac{\omega y^2}{2}}\ , \ E_{n,m}=\ \omega[2(n+m)+p_2+%
\frac{3}{2}]\ ,
\end{equation}
where $n,m = 0,1,2, \ldots$; the parameter $p_2$ is defined by the relation $%
B = p_2(p_2-1)$. The degree of degeneracy is given by the number of
partitions of a nonnegative integer into the sum of two nonnegative integers.

We perform a gauge rotation of $H_{II} (x,y)$ with the ground state
eigenfunction (\ref{HII.2}), $\Psi_{0,0}(x,y)$ as a gauge factor and then a
change of variables
\[
h^{II} \equiv \frac{1}{\omega}( \Psi_{0,0}(x,y))^{-1} H_{II} (x,y)
\Psi_{0,0}(x,y) |_{t=\sqrt{2\omega} x, u=\omega y^2} =
\]
\begin{equation}  \label{HII.3}
-\partial^2_t - 2u\partial^2_u + 2t\partial_t + (2u - 1 -2p_2)\partial_u +%
\frac{3}{2} + p_2\ .
\end{equation}
It is easy to check that the operators
\begin{equation}
{\hat x}_C^{II}= 2\partial^2_t - 4u\partial^2_u - 4t\partial_t + 2(2u - 1
-2p_2)\partial_u -1 + 2p_2\ ,
\end{equation}

\begin{equation}
{\hat{x}}_{P}^{II}\ =-4tu\partial _{u}^{2}+4u\partial
_{tu}^{2}-(2u-1-2p_{2})\partial _{t}-2t(1+2p_{2})\partial _{u}\ ,
\end{equation}
generate an infinite dimensional algebra. They obey the commutation
relations
\[
\lbrack h^{II},{\hat{x}}_{C}^{II}]\ =\ [h^{II},{\hat{x}}_{P}^{II}]\ =\ 0\ ,
\]
and
\[
\lbrack {\hat{x}}_{C}^{II},{\hat{x}}_{P}^{II}]=-32u\partial
_{tuu}^{3}-16(1+2p_{2}-2u)\partial _{tu}^{2}+32tu\partial _{uu}^{2}+
\]
\
\begin{equation}
+8(1+2p_{2}-2u)\partial _{t}+16t(1+2p_{2})\partial _{u}]\ .
\end{equation}

The operators $h^{II},{\hat{x}}_{C}^{II},{\hat{x}}_{P}^{II}$ as well as the
commutator $[{\hat{x}}_{C}^{II},{\hat{x}}_{P}^{II}]$ can be immediately
rewritten in terms of the generators (\ref{sl3}) and have the form
\begin{equation}
h^{II}=-J_{1}J_{1}-2J_{4}J_{2}+2J_{3}+2J_{4}-(1+2p_{2})J_{2}+\frac{3}{2}%
+p_{2}\ ,  \label{HII.i1}
\end{equation}
\begin{equation}
{\hat{x}}%
_{C}^{II}=2J_{1}J_{1}-4J_{4}J_{2}-4J_{3}+4J_{4}-2(1+2p_{2})J_{2}-1+2p_{2}\ ,
\label{HII.i2}
\end{equation}

\begin{equation}
{\hat{x}}_{P}^{II}\ =\
-4J_{4}J_{6}+4J_{1}J_{4}-2J_{5}+(1+2p_{2})J_{1}-2(1+2p_{2})J_{6}\ ,
\label{HII.i3}
\end{equation}

Evidently, the operators (\ref{HII.i1}),(\ref{HII.i2}),(\ref{HII.i3})
preserve the same triangular flag of polynomials as in the Case I but in
variables $x,u$:
\[
h(x,u): \mathcal{P}_{n}(x,u)\mapsto \mathcal{P}_{n}(x,u)
\]
where $\mathcal{P}_{n}(x,u)=\langle x^p u^q|0\leq p+q \leq n\rangle$. Thus,
the operators $h^{II},{\hat x}_C^{II},{\hat x}_P^{II}$ have infinitely many
finite dimensional invariant subspaces and infinitely many polynomial
eigenfunctions.

\textbf{Case III.}

The third superintegrable Hamiltonian
\begin{equation}  \label{HIII.1}
H_{III}(x,y) = - \frac{1}{2}(\partial^2_x + \partial^2_y) + \frac{\alpha}{2r}%
+\frac{1}{4r^2} \bigg(\frac{\beta_1}{\cos^2 \frac{\phi}{2}}+\frac{\beta_2}{%
\sin^2 \frac{\phi}{2}}\bigg)\ ,
\end{equation}
where $\beta_{1,2} > - 1/8$ are parameters and $x=r\cos{\phi}, y=r\sin{\phi}$%
. It admits the separation of variables in polar and parabolic coordinates.
In parabolic coordinates it has the form
\begin{equation}  \label{HIII.2}
H_{III}(\xi,\eta) = - \frac{1}{2} \frac{1}{\xi^2+\eta^2}(\partial^2_{\xi} +
\partial^2_{\eta}) + \frac{1}{\xi^2+\eta^2} \bigg(2\alpha+ \frac{\beta_1}{%
\xi^2}+ \frac{\beta_2}{\eta^2}\bigg)\ ,
\end{equation}
with $x=\frac{1}{2}(\xi^2-\eta^2)\, ,\, y=\xi \eta$. The eigenfunctions
corresponding to the energy $E$ are
\begin{equation}  \label{HIII.21}
\Psi_{n,m} = \xi^{p_1} \eta^{p_2} L^{(-1/2+p_1)}_n (\sqrt{-2 E} \xi^2)
L^{(-1/2+p_2)}_m(\sqrt{-2E} \eta^2) e^{-\sqrt{-E/2} (\xi^2+\eta^2)}
\end{equation}
where $2\beta_1=p_1(p_1-1), 2\beta_2=p_2(p_2-1)$. It is easy to see that the
Schroedinger equation $H_{III}\Psi=E\Psi$ can be transformed into
\begin{equation}  \label{HIII.3}
[ - \frac{1}{2}(\partial^2_{\xi} + \partial^2_{\eta}) -E(\xi^2+\eta^2)+
\frac{\beta_1}{\xi^2}+ \frac{\beta_2}{\eta^2}]\Psi= -2\alpha \Psi\ .
\end{equation}
We introduce the notation
\begin{equation}
Q_{III}\equiv (\xi^2+\eta^2)(H_{III}-E)-2\alpha \ .  \label{28}
\end{equation}
Equation (\ref{HIII.3}) can be written as
\begin{equation}
Q_{III} \Psi = -2\alpha \Psi \ ,  \label{29}
\end{equation}
and $Q^{III}$ can be related to $H^{I}$
\begin{equation}  \label{HIII.4}
Q_{III}= H_I(\xi,\eta;-E,\beta_1,\beta_2)\ ,
\end{equation}
(cf. (\ref{H.1})). We draw the striking conclusion that the Hamiltonian of
the first problem (Case I) written in \textit{Cartesian} coordinates
coincides with a modified third Hamiltonian $Q^{III}$ written in \textit{%
parabolic} coordinates (!). The parameter $(-2\alpha)$ plays the role of
spectral parameter which is the energy in Case I, $(-2\alpha)
\longleftrightarrow E^{I} $. Thus, the analysis performed for the Case I can
be repeated for this case.

We perform a gauge rotation of the operator $Q_{III}$ with a gauge factor
given by the multiplier figuring in eq.(\ref{HIII.21}),
\begin{equation}
M_{III}=\xi^{p_1}\eta^{p_2}e^{-\sqrt{-E/2}(\xi^2+\eta^2)}\ .
\end{equation}
Notice that in this case, contrary to those of $H_{I}$ and $H_{II}$, the
gauge factor is not universal. It depends on the energy $E$. Thus $E$ in the
multiplier $M$ is the considered energy, not the ground state one.

Thus we have
\[
q^{III} \equiv \frac{1}{\sqrt{-2E}} M_{III}^{-1} Q_{III} (x,y) M_{III}\ ,
\]
and with a change of coordinates $t=\sqrt{-2E}\xi^2, u=\sqrt{-2E}\eta^2$ we
get
\begin{equation}  \label{qIII}
q^{III}= -2t\partial^2_t - 2u\partial^2_u + 2t\partial_t + 2u \partial_u -
(2p_1+1) \partial_t - (2p_2+1) \partial_u +1+p_1+p_2 \ .
\end{equation}
This operator coincides \textit{exactly} with the operator $h^{I}$ (\ref
{HI.1}). Its spectrum is equal to $-2\alpha /\sqrt{-E} = 2(n+m)+1+p_1+p_2 \
,\ n,m=0,1,2,\ldots$.

Thus, the operator $q^{III}$ commutes with ${\hat x}_C^I,{\hat x}_R^I$. We
shall call these operators ${\hat x}_C^{III},{\hat x}_R^{III}$. The operator
$q^{III}$ can be rewritten in terms of the generators $\{ J_{1,...,6}\}$ of
the maximal parabolic subalgebra of $sl(3,\mathbb{R})$ (see (\ref{sl3})). We
see that the Hamiltonian (\ref{HIII.1}) is exactly solvable.

The above observation gives rise to an interesting question about the
connection between the operators commuting with original Hamiltonian (\ref
{HIII.1}) and the operators ${\hat{x}}_{C}^{III},{\hat{x}}_{R}^{III}$.

The operators that commute with $Q_{III}$ of eq. (\ref{28}) can be read off
from those of case $H_{I}$.  They are in parabolic coordinates
\[
X_{C}^{III}=-\frac{1}{2}(\partial _{\xi }^{2}-\partial _{\eta }^{2})-E(\xi
^{2}-\eta ^{2})+\frac{\beta _{1}}{\xi ^{2}}-\frac{\beta _{2}}{\eta ^{2}},
\]
\begin{equation}
X_{R}^{III}=(\xi \partial _{\eta }-\eta \partial _{\xi })^{2}-2(\xi
^{2}+\eta ^{2})(\frac{\beta _{1}}{\xi ^{2}}+\frac{\beta _{2}}{\eta ^{2}}).
\label{XIII}
\end{equation}
The operators that commute with the original Hamiltonian $H_{III}$ of eq.
(25), if written in parabolic coordinates, are \cite{Fris:1965,
Wint:1966,Sheftel:2000}
\[
X_{P}=\frac{1}{\xi ^{2}+\eta ^{2}}(\eta ^{2}\partial _{\xi }^{2}-\xi
^{2}\partial _{\eta }^{2}+2\alpha (\xi ^{2}-\eta ^{2})-2\beta _{1}\frac{\eta
^{2}}{\xi ^{2}}+2\beta _{2}\frac{\xi ^{2}}{\eta ^{2}}),
\]
\begin{equation}
X_{R}=(\xi \partial _{\eta }-\eta \partial _{\xi })^{2}-2(\xi ^{2}+\eta
^{2})(\frac{\beta _{1}}{\xi ^{2}}+\frac{\beta _{2}}{\eta ^{2}}).
\end{equation}

Let us consider, quite generally a Hamiltonian $H$ and an operator $Q$,
defined by the relation
\begin{equation}
H=\frac{Q+K}{\xi ^{2}+\eta ^{2}}+E,
\end{equation}
where $K$ is a constant and $E$ is the energy. Let $X$ be an operator
commuting with the Hamiltonian: $\left[ H,X\right] =0.$ The operator $Q$
then satisfies
\begin{equation}
\left[ Q,X\right] =\left( \xi ^{2}+\eta ^{2}\right) \left[ X,\frac{1}{\xi
^{2}+\eta ^{2}}\right] \left( \xi ^{2}+\eta ^{2}\right) \left( H-E\right) .
\end{equation}
Thus, if the operator $X$ commutes with $H$ (strongly, as an operator),
it commutes weakly, on functions $\Psi $ satisfying $\left(
H-E\right) \Psi =0,$ with $Q$.
So, in order to relate operators that commute with $H$ and those that
commute with $Q$, we have to consider linear combinations of the type $%
AX_{P}+BX_{R}+f\left( \xi ,\eta \right) $ $\left( H-E\right) .$ Here $A$ and
$B$ are constants, but $f\left( \xi ,\eta \right) $ can be any function, since
$\ H-E$ vanishes on the \textit{''energy shell''}.

Let us return to the problem at hand, i.e. the system characterized by the
Hamiltonian $H_{III}$, or equivalently, by the operator $Q_{III}$.
We have the following simple relation between the original integrals
$X_{P}$ and $X_{R}$ and the modified ones (\ref{XIII})
\begin{equation}
X_{C}^{III}=-X_{P}+\left( \xi ^{2}-\eta ^{2}\right) \left( H-E\right)
,\qquad X_{R}^{III}=X_{R}.  \label{XCIII}
\end{equation}

\textbf{Case IV.}

The fourth superintegrable Hamiltonian admits the separation of variables in
two mutually perpendicular parabolic systems of coordinates. In the usual
parabolic coordinates $\xi ,\eta $ it has the form
\begin{equation}
H_{IV}(\xi ,\eta )=-\frac{1}{2}\frac{1}{\xi ^{2}+\eta ^{2}}(\partial _{\xi
}^{2}+\partial _{\eta }^{2})+\frac{1}{\xi ^{2}+\eta ^{2}}\bigg(2\alpha
+\beta \xi +\gamma \eta \bigg)\ .  \label{HIV.1}
\end{equation}
The eigenfunctions are products of Laguerre polynomials times an energy
dependent multiplier, namely
\[
M_{IV}=(\xi -\beta /{2E})^{p_{1}}(\eta -\gamma /{2E})^{p_{2}}e^{-\sqrt{-E/2}[%
(\xi -\beta /{2E})^{2}+(\eta -\gamma /{2E})^{2}]}
\]
with a condition $\alpha +\frac{\beta ^{2}}{4E}+\frac{\gamma ^{2}}{4E}+\sqrt{%
-2E}=0$ and $p_{1},p_{2}=0,1$.

The corresponding Schroedinger equation $H_{IV}\Psi=E\Psi$ can be
rewritten as
\begin{equation}  \label{HIV.3}
[ - \frac{1}{2}(\partial^2_{\xi} + \partial^2_{\eta}) - E(\xi-\frac{\beta}{2E%
})^2-E(\eta-\frac{\gamma}{2E})^2 ]\Psi= (-2\alpha -\frac{\beta^2+\gamma^2}{4E%
}) \Psi\ .
\end{equation}
For the operator on the left hand side we introduce the notation
\[
Q_{IV}\equiv (\xi^2+\eta^2)(H^{IV}-E)-2\alpha -\frac{\beta^2+\gamma^2}{4E} \
.
\]
Equation (\ref{HIV.3}) can be written as
\[
Q_{IV} \Psi = {\tilde \alpha} \Psi \ ,
\]
with the new spectral parameter ${\tilde \alpha}=-2\alpha -\frac{%
\beta^2+\gamma^2}{4E}$.

The operator $Q^{IV}$ can be related to $H^{I}$
\begin{equation}  \label{HIV.4}
Q_{IV}= H_I(\xi,\eta;-E,0,0)\ ,
\end{equation}
(cf. (\ref{H.1}), (\ref{HIII.4})). We see that similarly to the Case III the
Hamiltonian of Case I written in \textit{Cartesian} coordinates coincides
with a modified fourth Hamiltonian $Q_{IV}$ written in \textit{parabolic}
coordinates. The parameter $\tilde \alpha$ plays the role of a spectral
parameter which was the energy in the first case, $\tilde \alpha
\longleftrightarrow E^{I} $. Thus, the analysis performed for the Case I can
be again repeated for this case.

Writing the gauge rotated operator $Q_{IV}$ in new coordinates
 $$t=\sqrt{-2E}(\xi-\beta/{2E})^2,u=\sqrt{-2E}(\eta-\gamma/{2E})^2$$
we get
\[
q^{IV} \equiv \frac{1}{\sqrt{-2E}} M_{IV}^{-1} Q_{IV} (x,y) M_{IV}
\]
\[
= -2t\partial^2_t - 2u\partial^2_u + 2t\partial_t + 2u \partial_u - (2p_1+1)
\partial_t - (2p_2+1) \partial_u +1+p_1+p_2 \ ,
\]
This operator coincides \textit{exactly} with the operator $h^{I}$ (\ref
{HI.1}) and $q^{III}$ (\ref{qIII}). Its spectrum is equal to $%
2(n+m)+1+p_1+p_2 \ ,\ n,m=0,1,2,\ldots$.

Thus, the operator $q^{IV}$ commutes with ${\hat x}_C^I,{\hat
x}_R^I$. We shall call these operators ${\hat x}_C^{IV},{\hat
x}_R^{IV}$. The operator $q^{IV}$ (as well as ${\hat
x}_C^{IV},{\hat x}_R^{IV}$) can be rewritten in terms of the
generators $\{ J_{1,...,6}\}$ of the maximal parabolic subalgebra
of $sl(3,\mathbb{R})$ (see (\ref{sl3})) and hence is exactly
solvable.

As in the Case III one can give the connection between the operators
commuting with original Hamiltonian (\ref{HIV.1}) and the operators ${\hat{x}%
}_{C}^{IV},{\hat{x}}_{R}^{IV}$. The operators commuting with
$H_{IV}$ are \cite{Fris:1965, Wint:1966,Sheftel:2000}
\[
X_{1}=\frac{1}{2\left( \xi ^{2}+\eta ^{2}\right) }\left\{ \eta ^{2}\partial
_{\xi }^{2}-\xi ^{2}\partial _{\eta }^{2}+2\alpha \left( \xi ^{2}-\eta
^{2}\right) +2\xi \eta \left( \gamma \xi -\beta \eta \right) \right\}\  ,
\]
\begin{equation}
X_{2}=\frac{1}{\left( \xi ^{2}+\eta ^{2}\right) }\left\{ \xi \eta \left(
\partial _{\xi }^{2}+\partial _{\eta }^{2}\right) +\left( -\beta \eta
+\gamma \xi \right) \left( \xi ^{2}-\eta ^{2}\right) -4\alpha \xi \eta \right\}
-\partial _{\xi \eta }^{2}\  .  \label{IV1}
\end{equation}

Two {\it second order} differential operators commuting with $Q_{IV}$ of
eq. (\ref{HIV.4}) can be related to (\ref{IV1}) and are given by
\begin{equation}
 Q^{(1)} = -2X_{1}+\left( \xi ^{2}-\eta ^{2}\right) (H_{IV}-E)+%
 \frac{\gamma ^{2}-\beta ^{2}}{4E}\ ,
\end{equation}
\begin{equation}
Q^{(2)} = X_{2}+2\xi \eta \left( H_{IV}-E\right) -\frac{\beta
\gamma }{2E}\ .  \label{IV2}
\end{equation}
From this point of view the fourth superintegrable system is
particularly simple. Explicitly, we have a new ``Hamiltonian"
\begin{equation}
Q_{IV}=-\frac{1}{2}\left( \partial _{\xi }^{2}+\partial _{\eta }^{2}\right)
-E\left[ \left( \xi -\frac{\beta }{2E}\right) ^{2}+\left( \eta -\frac{\gamma
}{2E}\right) ^{2}\right] \ ,  \label{IV3}
\end{equation}
which corresponds to the harmonic oscillator, while one of the
commuting operators satisfies
\begin{equation}
Q^{(1)} = X^I_C \ ,
\end{equation}
where the integral $X^I_C$ (see (8)) is written in the coordinates
$(\xi-\beta/{2E}), (\eta-\gamma/{2E})$.

In turn, the operator $ \widehat{X}_{R}^{I}$ reduces to
\begin{equation}
 \widehat{X}_{R}^{I}= [( \xi -\frac{\be }{2E}) \pa_{\eta
 }- (\eta -\frac{\gamma }{2E}) \pa_{\xi} ]^2 \equiv L_z^2 \ .
\end{equation}
Thus, the superintegrable system characterized by the Hamiltonian
$H_{IV}(\xi,\eta)$ has been reduced to a harmonic oscillator with
''frequency'' $\omega =\sqrt{-E/2}$ and a displaced equilibrium
point $\xi =\beta /2E$, $\eta =\gamma /2E$. It is well-known that
the harmonic oscillator is invariant under an $SU(2)$ group.
Indeed, we find that the ''Hamiltonian'' $Q_{IV}$ commutes with
$Q^{(1,2)}$ and $L_z$. The operators (42), (\ref{IV2}),
(\ref{IV3}) and $L_z$ form the basis of a $u(2)$ symmetry algebra
with $Q_{IV}$ as its center.

\section{Conclusions}

In general, integrability of a quantum system does not guarantee that
spectrum and eigenfunctions can be found in sufficiently explicit form. The
simplest example of this situation is given by one-dimensional quantum
dynamics which is integrable for any potential. The main message of the
present work is that the superintegrable systems on $E_2$ with the integrals
given by second order differential operators are exactly solvable as well.
We conjecture that the property of exact solvability will remain valid for
higher dimensional superintegrable systems of the above mentioned type.

\section*{Acknowledgements}

The reaserch reported here was initiated during a visit of P.T. and A.T. to
the CRM, Universit\'{e} de Montr\'{e}al and finished during a visit of P.W.
to the Universit\`{a} di Lecce. We thank both institutions for their
hospitality. P.T. was supported in part by INFN, sezione di Lecce, A.T. was
supported in part by CONACYT grant 25427-E (Mexico), P.W. by a research
grant from NSERC of Canada.

\begingroup\raggedright

\endgroup

\end{document}